\documentclass[12pt]{article}

\usepackage{graphicx}
\usepackage[left=2.5cm,top=2.5cm,right=2.5cm,bottom=2.5cm,nohead,nofoot]{geometry}
\usepackage{setspace}
\usepackage{natbib}

\usepackage[ruled]{algorithm2e}
\usepackage{amsmath}
\usepackage{amssymb}
\usepackage{subfigure}
\usepackage[normalem]{ulem}

\newcommand{\x}[1]{#1}
\newcommand{\xo}[1]{}

\newcommand{\imloc}{}%figures/}

\SetAlFnt{\small}
\SetAlCapFnt{\small}
\SetAlCapNameFnt{\small}
\SetAlCapHSkip{0pt}
\IncMargin{-\parindent}

\title{\x{Contrasting Multiple Social Network Autocorrelations for Binary Outcomes, With Applications To Technology Adoption}}
\author{Bin Zhang \and A.C. Thomas \and Patrick Doreian \and David Krackhardt \and Ramayya Krishnan}
\date{\today}

\begin{document}

\maketitle

\begin{abstract}
The rise of socially targeted marketing suggests that decisions made by consumers can be predicted not only from their personal tastes and characteristics, but also from the decisions of people who are close to them in their networks. One obstacle to consider is that there may be several different measures for ``closeness'' that are appropriate, either through different types of friendships, or different functions of distance on one kind of friendship, \x{where only a subset of these networks may actually be relevant}. Another is that these decisions are often binary and more difficult to model with conventional approaches, both conceptually and computationally. \x{To address these issues, we present a hierarchical model for individual binary outcomes that uses and extends the machinery of the auto-probit method for binary data. We demonstrate the behavior of the parameters estimated by the multiple network-regime auto-probit model (m-NAP) under various sensitivity conditions, such as the impact of the prior distribution and the nature of the structure of the network, and demonstrate on several examples of correlated binary data in networks of interest to Information Systems, including the adoption of Caller Ring-Back Tones, whose use is governed by direct connection but explained by additional network topologies.}
\end{abstract}

%\category{C.2.2}{Computer-Communication Networks}{Network Protocols}

%\terms{Design, Algorithms, Performance}

%\keywords{Social network, autocorrelation model, diffusion, Bayesian method}

%\acmformat{Zhang, B., Thomas, A. C., Doreian, P., Krackhardt, D., Krishnan, R.  2012. A Model for the Analysis of Adoption Behavior through Network Effects.}

%\begin{bottomstuff}
%
%Author's addresses: B. Zhang, Heinz College,
%Carnegie Mellon University; A. C. Thomas, Department of Statistics, Carnegie Mellon University; Patrick Doreian, Department of Sociology, University of Pittsburgh; David Krackhardt  {and} Ramayya Krishnan, Heinz College, Carnegie Mellon University.
%\end{bottomstuff}

\onehalfspacing

\section{Introduction}
\x{The prevalence and widespread adoption of online social networks have made the analysis of these networks, particularly the behaviors of individuals embedded within, an important topic of study in information systems \cite{Agarwal08,Oinas10}, building off previous work in the context of technology diffusion \cite{Brancheau90,Chatterjee90,Premkumar94}.} While past investigations into behavior in networks were typically limited to hundreds of people, contemporary data collection and retrieval technologies enable easy access to network data on a much larger scale\xo{, potentially billions of nodes and trillions of ties}. Analyzing the behavior of these individuals, such as their purchasing or technology adoption tendencies, requires statistical techniques that can handle both the scope and the complexity of the data. \\

The social network aspect is one such complexity. Researchers once assumed that \x{an individual's decision to purchase a product or adopt a technology is solely associated with their personal attributes,} such as age, education, and income \cite{Kamakura89,Allenby98}, though this could be due both to a lack of social network data and a mechanism for handling it; indeed, recent developments have shown that their decisions are associated with the decisions of an individual's neighbors in their social networks \cite{Bernheim94,Manski00,Smith04}. This could be due to a ``contagious'' effect, where someone imitates the behavior of their friends, or an indication of latent homophily, in which some unobserved and shared trait drives \x{both} the tendency for two people to form a friendship and for each \xo{of them }to \x{adopt} \citep{aral2009distinguishing,shalizi2011hac}; either social property will increase the ability to predict a person's adoption behavior beyond their \x{personal} characteristics. \\

\x{Each of these produces} outcomes that \xo{, when viewed statically, }are correlated between members of the network who are connected. A popular approach to study this phenomenon is to use a model with explicit autocorrelation between individual outcomes, defined with a single network structure term. With the depth of data now available, an actor is very often observed to be a member of multiple distinct but overlapping networks, such as a friend network, a work colleague network, a family network, and so forth, and each of these networks may have some connection to the outcome of interest, so a model that condenses all networks into one relation will be insufficient. While models have been developed to include two or more network autocorrelation terms, such as \citet{doreian89}, these do not allow for the immediate and principled inclusion of binary outcomes; other methods to deal with binary outcomes on multiple networks, such as \citet{Yang03}, instead take a weighted average of other networks in the system, combining them into one, which has the side effect of constraining the sign of each network autocorrelation component to be identical, which may be undesirable if there are multiple effects thought to be in opposition to one another. \\

To deal with these issues, we construct a model for binary outcomes \x{that uses} the probit framework, \xo{which} allow\x{ing} us to represent these outcomes as if they are dichotomized outcomes from a multivariate Gaussian random variable; this is then presented as in \citet{doreian89} to have multiple regimes of network autocorrelation. We first use the Expectation-Maximization algorithm (EM) to find a maximum likelihood estimator for the model parameters, then use Markov Chain Monte Carlo, a method from Bayesian statistics, to develop an alternate estimate based on the posterior mean. We also study the sensitivity of both solutions to the change of parameters' prior distribution. Preliminary experiments show that the E-M solution to this model is degenerate, and cannot produce a usable variance-covariance matrix for parameter estimates, and so the MCMC method is preferred. Our software is also validated by using the posterior quantiles method of \citet{Cook06}. We ensure that the parameter estimates from the model are correct by testing first on simulated data, before moving on to real examples of network-correlated behavior. \\

The rest of the paper is organized as follows. We discuss the literature on the network \x{autocorrelation} model in Section \ref{sec:literature}. Our two estimation algorithms for the multi-network autoprobit, based on EM and MCMC, are presented in Section \ref{sec:method}. In Section \ref{sec:experi} we present the results of experiments for software validation and parameter estimation behavior observation. Conclusions and suggestions for future work complete the paper in Section \ref{sec:conclusion}.
%
%
%
%---------------------------------------------------------------
%---------------------------------------------------------------
\section{Background}\label{sec:literature}
\x{[[Previously: Literature]]}
Network models of behavior are developed to study the process of social influence on the diffusion of a behavior, which is the process ``by which an innovation is communicated through certain channels over time among the members of a social system ... a special type of communication concerned with the spread of messages that are perceived as new ideas'' \cite{Rogers62}. These models have been widely used to study diffusion since the \cite{Bass69} model, \xo{which is} a population-level approach that assumes that everyone in the social network has the same probability of interacting. Such assumption is not realistic because given a large social network, the probability of any random two nodes connecting to each other is not the same; for example, people with closer physical distance communicate more and are likely to exert greater influence on each other. A refinement to this approach is a model where the outcomes of neighboring individuals are explicitly linked, such as the simultaneous autoregressive model (SAR). The general method of SAR is described in \citet{Anselin88} and \citet{Cressie93}; it considers simultaneous autoregression on the residuals of the form
\begin{align}
%\[
\mathbf{y} = \mathbf{X}\boldsymbol{\beta}+ \boldsymbol{\theta},\ \boldsymbol{\theta} = \rho \mathbf{W} \boldsymbol{\theta} + \boldsymbol{\epsilon} \nonumber
%\]
\end{align}
\noindent where $\mathbf{y}$ is a vector of observed outcomes, in this case consumer choice; $\mathbf{X}$ is a vector of explanatory variables. Rather than an independent error term, $\boldsymbol{\theta}$ represents error terms whose correlation is specified by $\mathbf{W}$, the social network matrix of interest, and $\rho$, the corresponding network autocorrelation, distributing a Gaussian error term $\epsilon_{i}$.

Maximum likelihood estimate solutions are provided by \citet{Ord75}, \citet{Doreian80,Doreian82}, and \citet{Smirnov05}. \\

\x{Standard} network \x{autocorrelation} models can only accommodate one network, \x{such as those of \citet{Burt87} and \citet{Leenders97}}. However, an actor is very often under influence of multiple networks, such as that of friends and that of colleagues. So if a research requires investigation of which \x{autocorrelation term} out of multiple networks plays the most significant role in consumers' decision, none of these models are adequate, and a model that can accommodate two or more networks is necessary.\\

%
%More here -- explain cohesion and structural equivalence more.
%Also: the ``same zip code'' model suggests that they may quite easily be confounded, to a degree; I know we get separation due to count data, but that's only across groups, not within.

Cohesion and structural equivalence are two competing social network models to explain diffusion of innovation. In the cohesion model, a focal person's adoption is influenced by his/her neighbors in the network. In the structural equivalence model, a focal person's adoption is influenced by the people who have the same position in the social network\x{, such as sharing many common neighbors}. While considerable work has been done on these models on real data, the question of which network model best explains diffusion has not been resolved. To approach this, \citet{doreian89} introduced \x{a model for} ``two regimes of network effects autocorrelation''\footnote{\x{The term ``network effects'' can refer to two directly related concepts: the autocorrelation between individual behaviors on a network, and the increased impact of a technology to an individual when used by more people within a network. Our meaning is the first, though we use the term {\em partial network autocorrelation} to avoid ambiguity.}} for continuous outcomes. \xo{Such a method allows us to investigate \x{influences} of two networks \xo{effect} on consumers' choices, so long as these choices reflect the type of data required.} The model is described as below: %
%
%\begin{align}
\[ \mathbf{y} = \mathbf{X} \boldsymbol{\beta} + \rho_{1}\mathbf{W}_{1}\mathbf{y} + \rho_{2}\mathbf{W}_{2}\mathbf{y} + \boldsymbol{\epsilon} \nonumber \]
%\end{align}
%
\noindent where $\mathbf{y}$ is the dependent variable; $\mathbf{X}$ is a vector of explanatory variables; each $\mathbf{W}$ represents a social structure underlying each autoregressive regime. \x{This model takes both interdependence of actors and their attributes, such as demographics, into consideration; these interdependencies are each described by a weight matrix $\mathbf{W}_i$. Doreian's model can capture both actor's intrinsic opinion and influence from alters in his social network.} \\

As this model \x{takes} a continuous dependent variable, \citet{Fujimoto11} \x{present} a plausible solution for binary outcomes by directly inserting an autocorrelation term $\mathbf{Wy}$ into the right hand side of a logistic regression:
\begin{align}
y_i & \sim \operatorname{Be}(p_i) \nonumber \\
\log(\frac{p_i}{1-p_i}) & = \mathbf{X} \boldsymbol{\beta} + \rho \sum_j \mathbf{W}_{ij}\mathbf{y}_{j} \nonumber
\end{align}
Due to its speed of implementation, this method is called ``quick and dirty" (QAD) by \citet{Doreian82}. Although it may support a binary dependent variable and multiple network terms, this model does not satisfy the assumption of logistic regression -- the observations are not conditionally independent, and the estimation results are biased. \x{\citet{Thomas2012SIM} shows that this method has more consequences than expected for the estimation procedure beyond simple bias; for example, in cases where $\mathbf{W}$ is a directed graph, networks that are directional cannot be distinguished from their reversed counterparts.}\\

\citet{Yang03} \x{propose} a hierarchical Bayesian autoregressive mixture model to analyze the effect of multiple network \x{autocorrelation terms} on a binary outcome. Their model can only technically accommodate one network effect, composed of several smaller networks that are weighted and added together. This model therefore assumes that all component network coefficients must have the same sign\footnote{\x{It is of course possible to specify terms in the $\mathbf{W}$ matrix as negative, to represent anticorrelation on a tie, but this must be done {\em a priori}, and is redundant in our approach.}}, and also be statistically significant or insignificant together. Such assumptions do not hold if the effect of any but not all of the component networks is statistically insignificant\x{, or of the opposite sign to the other networks, so a method that estimates coefficients for each $\mathbf{W}$ separately is necessary for our applications. We contrast our method with the Yang-Allenby grand $\mathbf{W}$ construction method, a finite mixture of coefficient matrices, in Appendix \ref{apx:W}.}
%
%
%
%---------------------------------------------------------------
%---------------------------------------------------------------
\section{Method}\label{sec:method}
We propose a variant of the auto-probit model that accommodates multiple regimes of network \x{autocorrelation terms} for the same group of actors, which we call the multiple network auto-probit model (m-NAP). We then provide two methods to obtain estimates for our model. The first is the use of Expectation-Maximization, which employs a maximum likelihood approach, and the second one is a Markov Chain Monte Carlo routine that treats the model as Bayesian. Detailed descriptions of both estimations are shown in Appendix \ref{apx:EM} and \ref{apx:MCMC}.
\subsection{Model Specification}
The actors are assumed to have \x{$k$} different types of network connections between them, where $\mathbf{W}_i$ is the $i^{th}$ network in question \x{$i \in \{1,...,k\}$}. $\mathbf{y}$ is the vector of \x{length $n$ of} observed binary choices, and is an indicator function of the latent preference of consumers $\mathbf{z}$. If $\mathbf{z}$ is larger than a threshold 0, consumers choose $\mathbf{y}$ as 1; if $\mathbf{z}$ is smaller than 0, then consumers would choose $\mathbf{y}$ as 0.
\begin{align}
 \mathbf{y} &= \mathbb{I}(\mathbf{z}>0)  \nonumber \\
\mathbf{z} &= \mathbf{X}\boldsymbol{\beta} + \boldsymbol{\theta} + \boldsymbol{\epsilon},\ \boldsymbol{\epsilon} \sim \operatorname{Normal}_{n}(0, I_{n}) \nonumber \\
\boldsymbol{\theta} &= \sum_{i=1}^{k} \rho_{i}\mathbf{W}_{i}\boldsymbol{\theta} + \mathbf{u}, \ \mathbf{u} \sim \operatorname{Normal}_{n}(0, \sigma^{2}I_{n}) \nonumber
\end{align}
$\mathbf{z}$ is a function of both exogenous covariates $\mathbf{X}$, autocorrelation term $\boldsymbol{\theta}$, \x{and individual error}. $\mathbf{X}$ is an $n\times m$ covariate matrix\xo{, such as $[1\ \mathbf{X}_{0}]$}\x{that includes a constant as its first column}; these covariates could be the exogenous characteristics of consumers. $\boldsymbol{\beta}$ is an $m\times 1$ coefficient vector associated with $\mathbf{X}$. $\boldsymbol{\theta}$ is the autocorrelation term, which is responsible for those nonzero covariances in the $\mathbf{z}$. $\boldsymbol{\theta}$ can be described as the aggregation of multiple network structure $\mathbf{W}_{i}$ and coefficient $\rho_{i}$. \x{Each} $\mathbf{W}_i$ \x{is a} network structure describing connections and relationships among consumers. \\

\x{Our model explicitly allows multiple competing networks that can be defined by different mechanisms on an existing basis of network ties; for example, $\mathbf{W}_{1}$ describes an effect acting directly on a declared tie, such as homophily or social influence, whereas $\mathbf{W}_{2}$ describes \x{the} structural equivalence due to those ties. It can also be that each $\mathbf{W}_i$ is defined by a different type of network edge, such as friendship, colleagueship, or mutual group membership; note that none of these relationships must be mutually exclusive. Each} coefficient $\rho_{i}$ describes the effect size of its corresponding network $\mathbf{W}_{i}$,\xo{. By accommodating multiple networks in an auto-probit model}\x{so that we can compare the relative scales of} competing network structures for the same group of actors embedded in social networks.\\

The error \x{term for} the model is modeled as \x{an augmented expression that} consists of two parts, $\boldsymbol{\epsilon}$ and $\mathbf{u}$. $\boldsymbol{\epsilon}$ is the unobservable error term of $\mathbf{z}$ \x{that describes individual-level variation that is not shared on the network,} and $\mathbf{u}$ is the error \x{that is then distributed along each network, accounting for the non-zero covariance between units.} If we marginalize \x{this model by integrating out} $\boldsymbol{\theta}$, all the unobserved interdependency will be isolated \x{in a single expression for the distribution of $\mathbf{z}$}, given parameters $\boldsymbol{\beta}$, $\rho$ and $\sigma^{2}$, \x{as multivariate} with mean $\mathbf{X}\boldsymbol{\beta}$ and variance $\mathbf{Q}$.
\begin{align}
\mathbf{z} &\sim \operatorname{Normal} \left(\mathbf{X}\boldsymbol{\beta}, \mathbf{Q} \right) \nonumber
\end{align}
where

\[ \mathbf{Q} = I_{n} + \sigma^{2} \left( I_{n} - \displaystyle{\sum_{i=1}^{k}} \rho_{i} \mathbf{W}_{i} \right)^{-1}\left(\left( I_{n} - \displaystyle{\sum_{i=1}^{k}} \rho_{i} \mathbf{W}_{i} \right)^{-1}\right)^{\top}. \]
\x{The non-standard form of the covariance matrix can therefore pose a significant computational issue.}
%
%
%---------------------------------------------------------------
%---------------------------------------------------------------
\subsection{Expectation-Maximization Solution}
We first develop an approach by maximizing the likelihood of the model using E-M. Since $\mathbf{z}$ is latent, we treat it as unobservable data, for which the E-M algorithm is one of the most used methods. Detailed description of our solution for $k$ regimes of network \x{autocorrelation} is in Appendix \ref{apx:EM}. \\

The method consists of two steps: first, estimate the expected value of functions of the unobserved $\mathbf{z}$ given the current parameter set $\boldsymbol{\phi}$, $(\boldsymbol{\phi}=\{\boldsymbol{\beta}, \boldsymbol{\rho}, \sigma^{2}\})$. Second, use these estimates to form a complete data set $\{\mathbf{y},\mathbf{X},\mathbf{z}\}$, with which we estimate a new $\boldsymbol{\phi}$ by maximizing the expectation of the likelihood of the complete data. \\

We first initialize the parameters \xo{need} to be estimated,
\begin{align}
\beta_{i} &\sim \operatorname{Normal}(\nu_{\beta},\Omega_{\beta}); \nonumber \\
\rho_{j} &\sim \operatorname{Normal}(\nu_{\rho},\Omega_{\rho}); \nonumber \\
\sigma^{2} &\sim \operatorname{Gamma}(a,b) \nonumber
\end{align}
where $i=1,...,m$, and $j=1,...,k$. Let these values equal $\boldsymbol{\phi}^{(0)}$.\\

\x{For the E-step, we} calculate the conditional expectation of \x{the log-likelihood, with respect to the augmented data,}
\begin{align}
G(\boldsymbol{\phi}\mid\boldsymbol{\phi}^{(t)}) &= \operatorname{E}_{\mathbf{z}\|\mathbf{y},\boldsymbol{\phi}^{(t)} } [\operatorname{log} L(\boldsymbol{\phi}\mid\mathbf{z,y})] \nonumber \\
&= -\frac{n}{2}\operatorname{log}2\pi - \frac{n}{2} \operatorname{log}\mid\mathbf{Q}\mid - \frac{1}{2} \sum_{i=1}^{n} \sum_{j=1}^{n} \check{q}_{ij}(\operatorname{E}[z_{i}z_{j}]-\operatorname{E}[z_{i}]X_{j}\beta - \operatorname{E}[z_{j}]X_{i}\beta + X_{i}X_{j}\beta^{2})  \nonumber
\end{align}
where $t$ is the \x{current step number} \xo{, $\mathbf{Q}=\operatorname{Var}(\mathbf{z})$, $\mathbf{Q} = I_{n} + \sigma^{2} \left( I_{n} - \sum_{i=1}^{k} \rho_{i} \mathbf{W}_{i} \right)^{-1}\left(\left( I_{n} - \sum_{i=1}^{k} \rho_{i} \mathbf{W}_{i} \right)^{-1}\right)^{\top}$,}
and $\check{q}_{ij}$ is \x{element $(i,j)$} in the matrix $\mathbf{Q}^{-1}$. \\

In the M-step, we maximize $\x{G}(\boldsymbol{\phi} \mid \boldsymbol{\phi}^{(t)})$ to get $\boldsymbol{\beta}^{t+1}$, $\boldsymbol{\rho}^{t+1}$ and $[\sigma^2]^{(t+1)}$ for the next step.
\x{\begin{align}
\boldsymbol{\beta}^{(t+1)} &= \underset{\boldsymbol{\boldsymbol{\beta}}}{\operatorname{arg\,max}}\ G(\boldsymbol{\beta} \mid \boldsymbol{\rho}^{(t)}, \boldsymbol{[\sigma^2]}^{(t)} ); \nonumber \\
\boldsymbol{\rho}^{(t+1)} &= \underset{\boldsymbol{\rho}}{\operatorname{arg\,max}}\ G(\boldsymbol{\rho} \mid \boldsymbol{\beta}^{(t+1)}, \boldsymbol{[\sigma^2]}^{(t)}  ); \nonumber \\
\boldsymbol{[\sigma^2]}^{(t+1)} &= \underset{[\sigma^2]}{\operatorname{arg\,max}}\ G(\boldsymbol{[\sigma^2]} \mid \boldsymbol{\beta}^{(t+1)}, \boldsymbol{\rho}^{(t+1)} ) \nonumber
\end{align}}
We replace $\boldsymbol{\phi}^{(t)}$ with $\boldsymbol{\phi}^{(t+1)}$ and repeat the E-step and M-step until all the parameters converge. Parameter estimates from the E-M algorithm converge to the MLE estimates \cite{Wu83}. \\

It is worth noting that the analytical solution for all the parameters is \x{not always possible. Consider the maximization with respect to the autocorrelation variance parameter $\sigma^{2}$:}
\begin{align}
[\sigma^2]^{(t+1)} &= \underset{[\sigma^2]}{\operatorname{arg\,max}}\ G(\boldsymbol{\phi} \mid\boldsymbol{\phi}^{(t)} ) \nonumber \\
\frac{\partial \operatorname{log}L}{\partial [\sigma^2]} &= \frac{\partial }{\partial [\sigma^2]} \left( - \frac{1}{2}\operatorname{log}\mid \mathbf{Q}\mid - \displaystyle{ \frac{1}{2} } (\mathbf{z}-\mathbf{X}\boldsymbol{\beta})^{\top}\mathbf{Q}^{-1}(\mathbf{z}-\mathbf{X}\boldsymbol{\beta}) \right) \label{aln:dLds2}
\end{align}
The first term at the the right hand side of Equation (\ref{aln:dLds2}) is:
\begin{align}
\frac{\partial }{\partial [\sigma^2]} \operatorname{log}\mid\mathbf{Q}\mid &= \frac{\partial }{\partial [\sigma^2]} \operatorname{log} \left\vert I_{n} + [\sigma^2] \left( I_{n} - \sum_{i=1}^{k} \rho_{i} \mathbf{W}_{i} \right)^{-1}\left(\left( I_{n} - \sum_{i=1}^{k} \rho_{i} \mathbf{W}_{i} \right)^{-1}\right)^{\top} \right\vert \nonumber
\end{align}
The second term is:
\begin{align}
& \frac{\partial }{\partial [\sigma^2]}(\mathbf{z}-\mathbf{X}\boldsymbol{\beta})^{\top}\mathbf{Q}^{-1}(\mathbf{z}-\mathbf{X}\boldsymbol{\beta}) \nonumber \\
&= \frac{\partial }{\partial [\sigma^2]} (\mathbf{z}-\mathbf{X}\boldsymbol{\beta})^{\top} \left( I_{n} + [\sigma^2] \left( I_{n} - \sum_{i=1}^{k} \rho_{i} W_{i} \right)^{-1}\left(\left( I_{n} - \sum_{i=1}^{k} \rho_{i} W_{i} \right)^{-1}\right)^{\top} \right)^{-1} (\mathbf{z}-\mathbf{X}\boldsymbol{\beta}) \nonumber
\end{align}
This is not solvable analytically, and numerical methods are needed to get the estimators for this parameter and for $\rho$. \\

\begin{figure}[htbp]
  \begin{center}
    \includegraphics[scale=.35]{\imloc 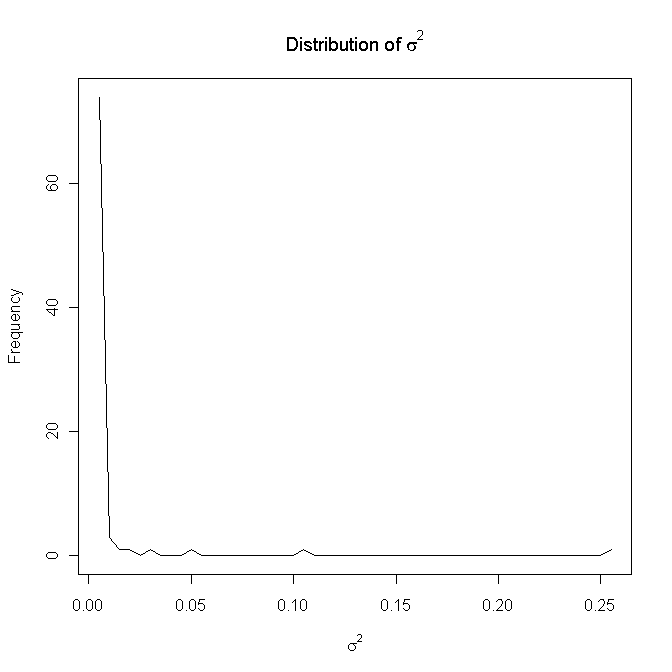}
    \caption{\x{An estimated probability distribution for $\sigma^{2}$, variance of $\boldsymbol{\theta}$. Maximum likelihood methods, such as the Expectation-Maximization method, will choose $\sigma^{2}=0$, a degenerate solution.}}\label{fig:s_dist}
  \end{center}
\end{figure}

As it happens, \xo{in its current form }the E-M algorithm produces a degenerate solution. This is because \x{it estimates} the mode of $\sigma^{2}$, the error term of the autocorrelation term $\theta$, \x{which} is at 0 (see Figure \ref{fig:s_dist}), \xo{so the estimated value of it by maximum likelihood is at 0, }and produces a singular variance-covariance matrix estimate using the Hessian approximation. Thus we have to find another solution.
\subsection{Full Bayesian Solution}
We \xo{then} turn to Bayesian methods. Since the observed choice of consumer's is decided by his/her unobserved preference, \x{this model} has a hierarchical structure, so it is natural to think of using a hierarchical Bayesian method. In addition to the model specification above, \xo{the} prior distributions for each of the highest-level parameters in the model are also \x{required}. \x{As before,} $\mathbf{y}$ is the observed dichotomous choice and calculated by the latent preference $\mathbf{z}$. \x{With Markov Chain Monte Carlo, we generate draws from a series of full conditional probability distributions}, \xo{which are} derived from the joint distribution. \x{We summarize the forms of the} full conditional distributions of all the parameters \xo{we need} to estimate in Table \ref{tbl:CCS}\x{, and in full in Appendix \ref{apx:MCMC}.} \\

%
%\begin{table}[h]
%\begin{center}
%\begin{tabular}{cll}
%  \hline
%  \textbf{Parameter} & \textbf{Density} & \textbf{Draw Type}\\
%  \hline
%  \hline
%  $\mathbf{z}$ & $\operatorname{TrunNormal}_{n}(\mathbf{X}\boldsymbol{\beta} + \boldsymbol{\theta}, I_{n})$ & Single \\
%  $\boldsymbol{\beta}$ & $\operatorname{Normal}_{n}(\boldsymbol{\nu}_{\beta}, \boldsymbol{\Omega}_{\beta}) $ & Parallel \\
%  $\boldsymbol{\theta}$ & $\operatorname{Normal}_{n}(\boldsymbol{\nu}_{\theta}, \boldsymbol{\Omega}_{\theta}) $ & Parallel \\
%  $\sigma^{2}$ & $\operatorname{InvGamma}(a, b) $ &  Parallel \\
%  $\rho_{i}$ & Metropolis step &  Sequential \\
%  \hline
%  \hline
%\end{tabular}
%\end{center}
%\caption{RCMS table for hierarchical Bayesian solution}
%\label{tbl:RCMS}
%\end{table}

% Table
\begin{table}%
\caption{\x{Cyclical conditional sampling steps for Markov Chain Monte Carlo}\label{tbl:CCS}}{%
\begin{tabular}{c|l|l}
\hline
  \textbf{Parameter} & \textbf{Density} & \textbf{Draw Type}\\
  \hline
  \hline
  $\mathbf{z}$ & $\operatorname{TrunNormal}_{n}(\mathbf{X}\boldsymbol{\beta} + \boldsymbol{\theta}, I_{n})$ & Parallel \\
  $\boldsymbol{\beta}$ & $\operatorname{Normal}_{n}(\boldsymbol{\nu}_{\beta}, \boldsymbol{\Omega}_{\beta}) $ & Parallel \\
  $\boldsymbol{\theta}$ & $\operatorname{Normal}_{n}(\boldsymbol{\nu}_{\theta}, \boldsymbol{\Omega}_{\theta}) $ & Parallel \\
  $\sigma^{2}$ & $\operatorname{InvGamma}(a, b) $ &  Single \\
  $\rho_{i}$ & Metropolis step &  Sequential \\
  \hline
  \hline
\end{tabular}}
\end{table}%
Given the observed choice of consumer, the latent variable $\mathbf{z}$ \x{is} generated from a truncated normal distribution with a mean of $\mathbf{X}\boldsymbol{\beta} + \boldsymbol{\theta}$ with unit error. The prior distributions of the parameters (shown in Table \ref{tbl:CCS} are \xo{generally adopted}adapted from \xo{the }priors proposed by \cite{Smith04}: \\

\begin{itemize}

\item \x{$\boldsymbol{\beta}$ follows a multivariate normal distribution with mean $\boldsymbol{\nu}_{\beta}$ and variance $\boldsymbol{\Omega}_{\beta}$. }

\item \x{$\sigma^{2}$ follows an inverse gamma distribution with parameters $a$ and $b$. }

\item \x{Each $\rho_i$ follows a normal distribution with mean $\nu_{\rho}$ and variance $\Omega_{\rho}$. }

\end{itemize}

% \xo{We then use Markov Chain Monte Carlo (MCMC) to generate draws of conditional posterior distributions for the parameters in 5 steps. Detailed description of my method, including the conditional distribution of all parameters, is given in Appendix \ref{apx:MCMC}.} \\
%
%

\x{The sampler algorithm was constructed in the R programming language, including a mechanism to generate data from the model. Validation of the algorithm was conducted using the method of posterior quantiles \citep{Cook06}, ensuring the correctness of the code for all analyses. Posterior quantiles is a simulation-based method that generates data from the model and verifies that the software can generate parameter estimate randomly around true parameter. For detailed description of the implementation, please see Appendix \ref{apx:quantile}.}

\subsection{Sensitivity to Prior Specification}

\begin{figure}[ht]
\centering
\subfigure[$\rho \sim \operatorname{Normal}(0,100)$]{
\includegraphics[scale=0.4]{\imloc 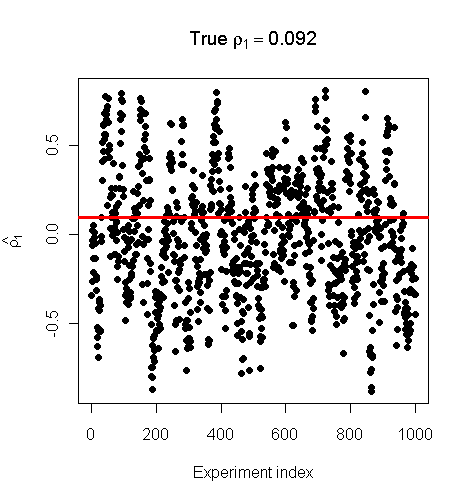}
\label{fig:rho1}
}
\subfigure[$\rho \sim \operatorname{Normal}(0.05,0.05^{2})$]{
\includegraphics[scale=0.4]{\imloc 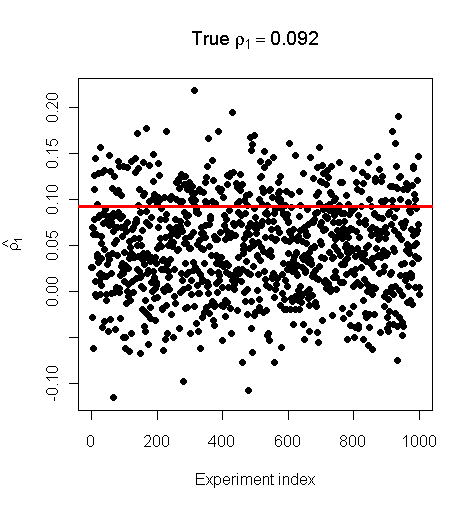}
\label{fig:rho2}
}
\label{fig:rhos}
\caption{\x{Testing the sensitivity of the inference of an autocorrelation parameter $\rho_1$ to the prior distribution. (a) The Markov Chain for a weakly informative prior distribution is consistent with the ``oracle'' value $\rho_1$, but the chain has significant temporal autocorrelation. (b) The Markov Chain with a strongly informative prior distribution has much less temporal autocorrelation, but is beholden to its prior distribution more than the data.}}
\end{figure}

We \xo{next} test the performance of the sampler using prior distributions that are closer to our chosen model than the trivial priors used to check the model code in order to assess the behavior of the algorithm under non-ideal conditions. \x{We demonstrate on data simulated from the model, using two pre-existing network configurations, and specify different prior distributions for each parameter.} \x{To demonstrate, we} choose a prior distribution for $\rho_1$ with high variance, $\rho \sim \operatorname{Normal}(0,100)$, . As shown in Figure \ref{fig:rho1}, the posterior draws of $\rho_1$ have \x{high temporal} autocorrelation. To compare, we choose a narrow prior distribution for $\rho_1$, $\rho_1 \sim \operatorname{Normal}(0.05,0.05^{2})$; the posterior draws for $\rho_1$ are shown in Figure \ref{fig:rho2}, and the \x{temporal} autocorrelation is considerably smaller\xo{, if not zero}. \x{With the volume of data under consideration, it is clear that the} posterior distribution of $\rho$ is sensitive to its prior distribution. \\ %In order to generate random estimates, we choose a narrow distribution, e.g. $\operatorname{Normal}(0.05,0.05^{2})$, as the prior for $\rho$.

\x{In most of our examples, we do not have a great deal of prior information available on any network parameters, suggesting that most of our analyses will be conducted with minimally informative prior distributions.} With such high autocorrelation between sequential draws, the effective sample size is extremely small. We therefore use a high degree of thinning to produce \x{a series of} uncorrelated draws from the posterior.\\

%Second, since $\rho$s are generated by using sequential draws, there is autocorrelation exist between consecutive draws, sometimes the autocorrelation is still high even the lag between two draws is large. The autocorrelation plot for $\rho_{1}$ (Figure \ref{fig:acf_r1}) shows two $\rho_{1}$ draws have strong correlation even the lag is larger than 300, so thinning for the draws is necessary.
%
%\begin{figure}[htbp]
%  \begin{center}
%    \includegraphics[scale=.4]{\imloc acf_r1.png}
%    \caption{Autocorrelation plot of $\rho$}
%  \end{center}
%  \label{fig:acf_r1}
%\end{figure}
%
%
\section{Applications}\label{sec:experi}

\subsection{Auto Purchase Data of Yang and Allenby (2003)}
We use \citeauthor{Yang03}'s \citeyear{Yang03} Japanese car data to \x{compare the findings of our method with those in the original study. The} data consists of \x{information on 857 purchase decisions of mid-size cars;} the dependent variable is whether \x{the car purchased was Japanese ($y_m=1$) or otherwise ($y_m=0$).} All the car models in the data are substitutable and have roughly similar prices. \\

\x{An important question of interest is} whether the preferences of Japanese car among \x{consumers} are interdependent or not. The interdependence in the network is measured by geographical location, where $W_{ij}=1$, if consumer $i$ and $j$ live in the same zip code, and 0, otherwise. Explanatory variables include actors' demographic information such as age, annual household income, ethnic group, education and other information such as the price of the car, whether the optional accessories are purchased for the car, latitude and longitude of the actor's location. \x{To construct a network, Yang and Allenby use whether the consumers' home address in the same zip code as the indicator of a connection. Thus the network structure $\mathbf{W}$, the cohesion, is the joint membership of same geographic area.}\\

\begin{figure}[htbp]
  \begin{center}
    \includegraphics[scale=.4]{\imloc 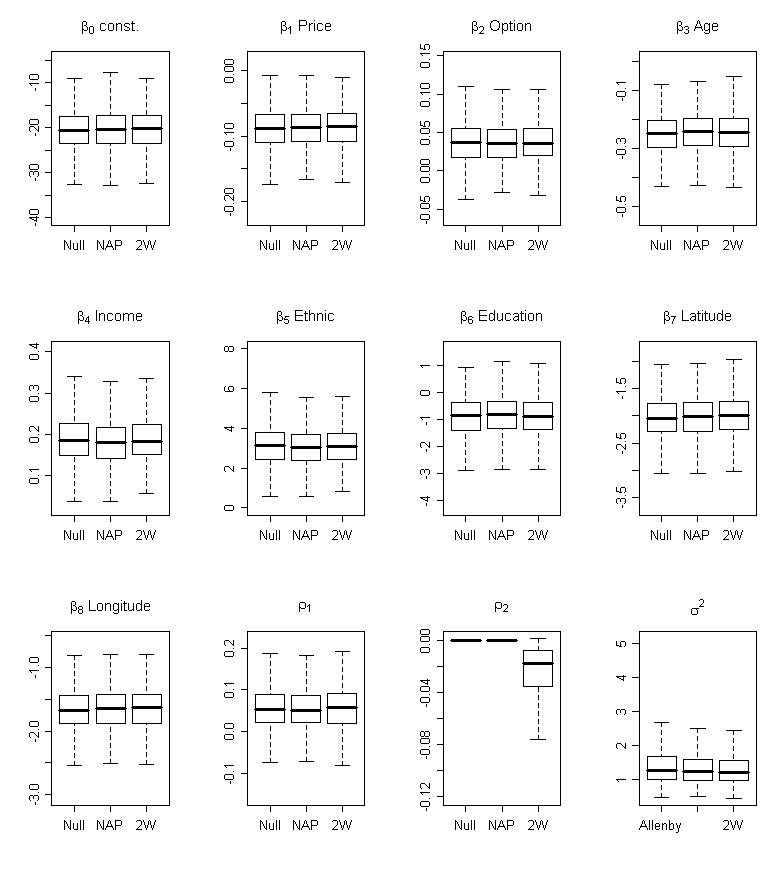}
    \caption{\x{A comparison of coefficient estimates between the Yang-Allenby method and m-NAP with 1 or 2 networks. The models give similar results, while noting that there is now a negative and statistically significant effect on the network representing structural equivalence. $\beta_{0}$: coefficient of constant term, $\beta_{1}$: coefficient of $\textbf{X}_{1}$, car price; $\beta_{2}$: coefficient of $\textbf{X}_{2}$, car's optional accessory; $\beta_{3}$: coefficient of $\textbf{X}_{3}$, consumer's age; $\beta_{4}$: coefficient of $\textbf{X}_{4}$, consumer's income; $\beta_{5}$: coefficient of $\textbf{X}_{5}$, consumer's ethnicity; $\beta_{6}$: coefficient of $\textbf{X}_{6}$, residence longitude; $\beta_{7}$: coefficient of $\textbf{X}_{7}$, residence latitude; $\rho_{1}$: coefficient of first network autocorrelation term, $\mathbf{W}_{1}$, cohesion; $\rho_{2}$: coefficient of the second network autocorrelation term, $\mathbf{W}_{2}$, structural equivalence; $\sigma^{2}$: estimated variance of the error term in autocorrelation.}}\label{fig:compare3m}
  \end{center}
\end{figure}

\x{By comparing the parameters of Yang and Allenby's model to those for m-NAP on the same dataset, with the same underlying definition of network structure, we contrast our approaches and demonstrate the value of separating the impact of various network autocorrelations.} The comparison of the coefficient estimates from Yang and Allenby and our Bayesian solution is shown in Figure \ref{fig:compare3m} \x{, for both explanatory variables and for network autocorrelations. We specify a second network term}
%In order to make a proper comparison, we set all the network \x{autocorrelations} except the first one as $\mathbf{0}_{n,n}$ matrix. Our $\mathbf{W}_{1}$ has the same definition as Yang and Allenby's $\mathbf{W}$. For the third method, we add one more network structure
$\mathbf{W}_{2}$ \x{to be the structural equivalence of two consumers, calculated as the simple adjacency distance between the two vectors representing individuals' connections to other individuals in the network to measure structural equivalence.} In a undirected network with non-weighted edges the adjacency distance between two nodes $i$ and $j$ is the \x{number of individuals who have different relationships to $i$ and $j$ respectively,}

\begin{align}\label{aln:eucli}
d_{ij} &= \sqrt{\sum_{k=1,k\neq i,j}^{N} (A_{ik}-A_{jk})^{2} },
\end{align}
where $A_{ik}=1$ if node $i$ and $k$ are neighbors, and 0 otherwise. The larger $d$ between node $i$ and $j$, the less structurally equivalent they are. We \x{use} the inverse of $d_{ij}$ plus one in order to construct a measure with a positive\x{, finite} relationship with role equivalence, so that ${s}_{ij} = \frac{1}{d_{ij}+1}$. \x{In our setting, a random element $A_{ij}$ in Equation (\ref{aln:eucli}) is from matrix $\mathbf{W}_{1}$, so $d_{ij}$ is the adjacency distance between any two vectors $\textbf{A}_{i}$ and $\textbf{A}_{j}$, representing consumer $i$'s connections, and consumer $j$'s connections to all the other consumers in the data, respectively. The inverse of $d_{ij}$ with an addition to 1 (to avoid zero as denominator), $s_{ij}$, becomes element of structural equivalence matrix $\mathbf{W}_{2}$.} \\

The comparison is shown in Figure \ref{fig:compare3m}. Each box contains the estimates of one parameter from three methods: \x{from left to right, Yang and Allenby, NAP with 1 network, and NAP with 2 networks}. All the coefficient estimates, $\hat{\boldsymbol{\beta}_{i}}$, $\hat{\boldsymbol{\rho}_{2}}$, and $\hat{\sigma}^{2}$ of the three methods have similar mean, standard deviation and credible interval. %Such results confirm again that NAP returns correct estimates of parameters in the model.
One thing interesting here is the effect size of the second network, structural equivalence, has a significant negative effect. \x{This} suggests a diminishing cluster effect; when the number of people in the cluster gets bigger, the influence does not increase proportionally. \xo{When the the structural equivalence between two customers is large, meaning they are in the same community (zip code), and the size, i.e. number of customers, of such community is large, so they have more common neighbors, thus more same scalar component in the vector.} \\

\subsection{Caller Ring-Back Tone Usage In A Mobile Network}

We \x{use m-NAP to} investigate \x{the purchase of} Caller Ring Back Tones (CRBT) within a \x{cellular phone} network, \x{a technology of increasing interest around the world}. \xo{CRBT is a special ring-back tone chosen by a subscriber, such as a song or a joke.}
When someone calls the subscriber of a CRBT, the caller does not hear the standard \xo{plain} ring-back tone but instead hears \x{a song, joke or other message chosen by the subscriber until the subscriber answers} the phone or the mailbox takes over. \xo{It has become more and more popular among cellular phone users globally.} \x{As soon as a CRBT is downloaded, it is set as the default ring back tone, and triggered automatically by all phone call. } Our data were obtained from a large Indian telecommunications company (source and raw data confidential). We have cellular phone call records and CRBT purchase records over a three-month period, and phone account holders' demographic information such as age and gender. \x{We extract a community of 597 users that are highly internally connected from a population with approximately 26 million unique users using the Transitive Clustering and Pruning (T-CLAP) algorithm \citep{zhang11}}. \xo{Those phone calls were initiated by customers both inside and outside of the company.} \x{Within this cluster, network edges are specified between users who call each other during the period of observation, as mutual} symmetric connection implies equal and stable \x{relationships} \citep{Hanneman05}, \x{rather than weaker relationships or calls related to businesses (inquiries or telemarketers).} \\
%Citation for T-CLAP
%

\x{We include several explanatory variables in this model:}

\begin{itemize}

\item The gender of the cellular phone account holder;

\item The age of the account holder;

\item The number of unique outbound connections from the user (known as the ``outdegree'').

\end{itemize}

\x{From our original network, we derive two matrices corresponding to cohesion and structural equivalence.} Cohesion assumes callers who make phone calls to each other will hear the called party's CRBT thus more likely to buy that ring-back tone or get interested in CRBT and eventually adopt the technology. Since the number of people a caller calls are drastically different, we normalize the cohesion matrix by dividing each row by the total number of adopters, to make the matrix element to be the percentage of adoption. Structural equivalence is \x{once again} defined as the adjacency distance between two callers. \xo{It measures how many common friends two callers share. The more common friends two callers share, the smaller the adjacency distance between them. However, in order to make the parameter of role equivalence have a positive relationship with high role equivalence, we use the inverse of role equivalence plus one instead.} \x{Here it is less clear that there is an obvious mechanism for how structural equivalence can impact adoption, as it relates to a relationship that does not expose the caller to the CRBT.} \\
\begin{figure}[htbp]
  \begin{center}
    \includegraphics[scale=.32]{\imloc 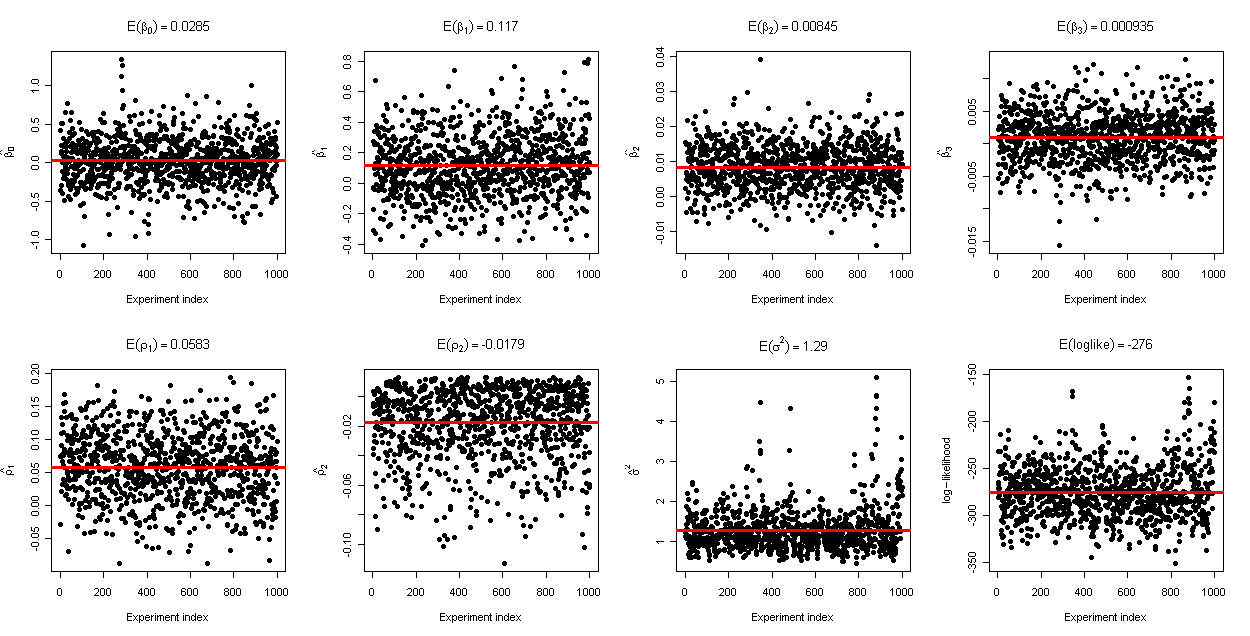}
    \caption{\x{Trace plot of CRBT network parameters. Description of parameters: $\beta_{0}$: coefficient of constant term; $\beta_{1}$: coefficient of consumer's gender; $\beta_{2}$: coefficient of consumer's age; $\beta_{3}$: coefficient of number of called contacts; $\rho_{1}$: coefficient of first network autocorrelation term, $\mathbf{W}_{1}$, cohesion; $\rho_{2}$: coefficient of the second network autocorrelation term, $\mathbf{W}_{2}$, structural equivalence; $\sigma^{2}$: estimated variance of the error term in autocorrelation; loglike: log-likelihood of $\mathbf{y}$. }}
  \end{center}
\end{figure}\label{fig:tracecrbt}

\x{We show estimates for each parameter of the model is shown in Figure \ref{fig:tracecrbt}.} Again, we observe a significant negative effect for structural equivalence. This new network autocorrelation, with a coefficient of opposite sign from that of the first network autocorrelation $\mathbf{W}_{1}$, cannot be identified by any earlier models. \xo{Our model is the first one that could possibly pick up additional autocorrelation term when single network autocorrelation could not account for all the interdependent influences among individuals embedded in the network.}

%---------------------------------------------------------------
%---------------------------------------------------------------
\section{Conclusion}\label{sec:conclusion}

We \x{have} introduced a new auto-probit model to study binary choice of a group of actors that have multiple network relationships among them. We specified the \x{fitting of the model for} both E-M and hierarchical Bayesian methods\xo{, and developed estimation solutions for both of them}. We found \x{that the} E-M solution cannot estimate the parameters \x{for this particular model}, thus only hierarchical Bayesian solution can be used here. We also validated our Bayesian solution by using \x{the} posterior quantiles method and the results show our software returns accurate estimates. Finally we compare the estimates returned by Yang and Allenby, NAP with one network effect (cohesion), and NAP with two network effects (cohesion and structural equivalence), by using real data. \xo{Experiments showed all three returned identical estimates, thus confirmed our software returns correct parameter estimates.} \\

%
%\xo{We intend to run our software on more benchmark data with better defined network structure. We also want to run more experiments with simulated populations to evaluate the properties of the solution. For example, let $\mathbf{W}$ have different features, such as network with randomly distributed edges, clustered edges, and skewed distributed edges etc.} \\

%
We want to ensure that the approach can recover variability in the network effect size. Assuming $\mathbf{W}\boldsymbol{\theta}$ has strong effect, we will vary $\rho$'s true value from small number to large number, and observe whether our solution can capture the variation. \\

\xo{Additionally, we want to compare our program with QAD, because although people know parameter estimates returned by \x{QAD} is biased, we do not know how different they are from the true value.} Finally we also want to study how multicollinearities between $\mathbf{X}$s, and between $\mathbf{X}$ and $\mathbf{W}\boldsymbol{\theta}$ affect estimated results.
%

% Bibliography
\bibliographystyle{ims}%acmsmall}
\bibliography{reference2}

% Appendix
\appendix
\section*{APPENDIX}
\setcounter{section}{1}

\subsection{E-M solution implementation}\label{apx:EM}
\subsubsection{Deduction}
First, get the distribution of $\boldsymbol{\theta}$.
\begin{align}
\left( I_{n} - \sum_{i=1}^{k} \rho_{i} \mathbf{W}_{i} \right)\boldsymbol{\theta} &= \mathbf{u} \nonumber \\
\boldsymbol{\theta} &= \left( I_{n} - \sum_{i=1}^{k} \rho_{i} \mathbf{W}_{i} \right)^{-1}\mathbf{u}  \nonumber \\
\boldsymbol{\theta} &\sim \operatorname{Normal} \left( 0, \sigma^{2} \left( I_{n} - \sum_{i=1}^{k} \rho_{i} \mathbf{W}_{i} \right)^{-1}\left(\left( I_{n} - \sum_{i=1}^{k} \rho_{i} \mathbf{W}_{i} \right)^{-1}\right)^{\top} \right)  \nonumber
\end{align}
Then get the distribution of $\mathbf{z}|\boldsymbol{\beta}, \boldsymbol{\rho}, \sigma^{2}$:
\begin{align}
\mathbf{z} &\sim \operatorname{Normal} \left(\mathbf{X}\boldsymbol{\beta}, \mathbf{Q} \right) \text{, where } \mathbf{Q} = I_{n} + \sigma^{2} \left( I_{n} - \sum_{i=1}^{k} \rho_{i} \mathbf{W}_{i} \right)^{-1}\left(\left( I_{n} - \sum_{i=1}^{k} \rho_{i} \mathbf{W}_{i} \right)^{-1}\right)^{\top} \nonumber
\end{align}
The joint distribution of $\mathbf{y}$ and $\mathbf{z}$ can transformed as:
\begin{align}
p(\mathbf{y}|\mathbf{z})p(\mathbf{z}|\boldsymbol{\beta}, \boldsymbol{\rho}, \sigma^{2}) &= p(\mathbf{y},\mathbf{z}| \boldsymbol{\beta}, \boldsymbol{\rho}, \sigma^{2}) \nonumber \\
 &= p(\mathbf{z}|\mathbf{y};\boldsymbol{\beta}, \boldsymbol{\rho}, \sigma^{2}) p(\mathbf{y}) \label{aln:PZY_PY}
\end{align}
The right side of equation (\ref{aln:PZY_PY}) are two distributions we already have, as shown below.
\begin{align}
p(\mathbf{y}) &= \displaystyle{ \frac{ \displaystyle{ \frac{1}{\sqrt{2\pi}} } \operatorname{exp}\left( \displaystyle{-\frac{1}{2}} (\mathbf{z}-\mathbf{X}\boldsymbol{\beta})^{\top}(\mathbf{z}-\mathbf{X}\boldsymbol{\beta}) \right) }{\Phi(\mathbf{X}\boldsymbol{\beta})}  } \mathbb{I}(\mathbf{z}>0) \nonumber \\
\mathbf{z}|\boldsymbol{\beta}, \boldsymbol{\rho}, \sigma^{2} & \sim \operatorname{Normal} (\mathbf{X}\boldsymbol{\beta}, \mathbf{Q}) \nonumber \\
\mathbf{z}|\mathbf{y}, \mathbf{X}; \boldsymbol{\beta}, \boldsymbol{\rho}, \sigma^{2} & \sim \operatorname{TrunNormal}(\mathbf{X}\boldsymbol{\beta}, \mathbf{Q}) \nonumber
\end{align}
Consider parameter $\boldsymbol{\beta}$ only,
\begin{align}
p(\boldsymbol{\beta},\mathbf{z}|\mathbf{y}) &= p(\boldsymbol{\beta}|\mathbf{z},\mathbf{y})p(\mathbf{z}|\mathbf{y}) \nonumber \\
\mathbf{z}|\mathbf{y},\mathbf{X};\boldsymbol{\beta} & \sim \operatorname{TrunNormal}(\mathbf{X}\boldsymbol{\beta},\mathbf{Q}) \nonumber
\end{align}
Assume $\operatorname{Var}(\mathbf{z}$)=1,
\begin{align}
L(\boldsymbol{\beta}|\mathbf{z}) &=  \frac{1}{\sqrt{2\pi}} \sum_{i=1}^{n} \operatorname{exp} \left( -\frac{1}{2} (z_{i}-X_{i}\beta)^{2}\right) \nonumber \\
\hat{\boldsymbol{\beta}} &= (\mathbf{X}^{\top}\mathbf{X})^{-1}\mathbf{X}^{\top} \mathbf{R}, \ \text{where } \mathbf{R}=\operatorname{E} [\mathbf{z}| \boldsymbol{\theta},\mathbf{y} ] \nonumber
\end{align}
Then include parameters, $\boldsymbol{\rho}$ and $\sigma^{2}$.
\begin{align}
\operatorname{E}[\mathbf{z}]^{(t+1)} &= \operatorname{E}[\mathbf{z}|\mathbf{y},\boldsymbol{\beta}^{(t)}] = f(\boldsymbol{\beta}^{(t)},\mathbf{y}) \nonumber \\
\operatorname{log}L(\boldsymbol{\beta},\boldsymbol{\rho},\sigma^{2}|\mathbf{z}) &= \operatorname{log}p(\mathbf{z}|\boldsymbol{\beta},\boldsymbol{\rho},\sigma^{2}) \nonumber \\
 &= \operatorname{log} \prod_{i=1}^{n} p(z_{i}|\boldsymbol{\beta},\boldsymbol{\rho},\sigma^{2}) \nonumber \\
 &= \displaystyle{ \sum_{i=1}^{n} } \operatorname{log}\displaystyle{\frac{1}{\sqrt{2\pi|\mathbf{Q}|}}} - \displaystyle{ \frac{1}{2} (\mathbf{z}-\mathbf{X}\boldsymbol{\beta})^{\top}\mathbf{Q}^{-1}(\mathbf{z}-\mathbf{X}\boldsymbol{\beta}) }  \nonumber \\
 &= \sum_{i=1}^{n}  \operatorname{log} \frac{1}{\sqrt{2\pi|\mathbf{Q}|}} - \left( \frac{1}{2} \mathbf{z}^{\top}\mathbf{Q}^{-1}\mathbf{z} - \mathbf{z}^{\top}\mathbf{Q}^{-1}\mathbf{X}\boldsymbol{\beta} - \mathbf{X}^{\top}\boldsymbol{\beta}\mathbf{Q}^{-1}\mathbf{z} + \mathbf{X}^{\top}\boldsymbol{\beta}\mathbf{Q}^{-1}\mathbf{X}\boldsymbol{\beta} \right) \label{aln:logL}
\end{align}
If decompose the matrices above as vector product, then:
\begin{align}
\text{(\ref{aln:logL})} &= \sum_{i=1}^{n}  \operatorname{log} \frac{1}{\sqrt{2\pi|\mathbf{Q}|}} - \frac{1}{2} \sum_{i=1}^{n} \sum_{j=1}^{n} (z_{i}-X_{i}\beta) \check{q}_{ij} (z_{j}-X_{j}\beta) \nonumber \\
 &= \sum_{i=1}^{n}  \operatorname{log} \frac{1}{\sqrt{2\pi|\mathbf{Q}|}} - \frac{1}{2} \sum_{i=1}^{n} \sum_{j=1}^{n} \check{q}_{ij}(z_{i}z_{j}-z_{i}X_{j}\beta - z_{j}X_{i}\beta + X_{i}X_{j}\beta^{2}) \nonumber
\end{align}
where $\check{q}_{ij}$ is the element in $\check{\mathbf{Q}}$, and $\check{\mathbf{Q}}=\mathbf{Q}^{-1}$.
\subsubsection{Expectation step}
In the expectation step, get the expected log-likelihood of parameters.
\begin{align}
Q(\boldsymbol{\phi}|\boldsymbol{\phi}^{(t)}) &= \operatorname{E}_{\mathbf{z}|\mathbf{y},\boldsymbol{\phi}^{(t)} } [\operatorname{log} L(\boldsymbol{\phi}|\mathbf{z,y})] \nonumber \\
&= \operatorname{E} \left[ \sum_{i=1}^{n} \operatorname{log} \frac{1}{\sqrt{2\pi|\mathbf{Q}}} \right] - \operatorname{E}\left[ \displaystyle{ \frac{1}{2} } (\mathbf{z}-\mathbf{X}\boldsymbol{\beta})^{\top}\mathbf{Q}^{-1}(\mathbf{z}-\mathbf{X}\boldsymbol{\beta}) \right] \nonumber \\
&= -\frac{n}{2}\operatorname{log}2\pi - \frac{n}{2} \operatorname{log}|\mathbf{Q}| - \frac{1}{2} \sum_{i=1}^{n} \sum_{j=1}^{n} \check{q}_{ij}(\operatorname{E}[z_{i}z_{j}]-\operatorname{E}[z_{i}]X_{j}\beta - \operatorname{E}[z_{j}]X_{i}\beta + X_{i}X_{j}\beta^{2})  \nonumber
\end{align}
where $\boldsymbol{\phi}$ is the parameter set, and $t$ is the number of steps.
\subsubsection{Maximization step}
In the maximization step, get the parameter estimates maximizing the expected log-likelihood. First, estimate $\boldsymbol{\beta}$
\begin{align}
\boldsymbol{\beta}^{(t+1)} &= \underset{\boldsymbol{\boldsymbol{\beta}}}{\operatorname{arg\,max}}\ Q(\boldsymbol{\phi} | \boldsymbol{\phi}^{(t)} ) \nonumber \\
 &= \underset{\boldsymbol{\boldsymbol{\beta}}}{\operatorname{arg\,max}}\ \displaystyle{ \sum_{i=1}^{n} } \operatorname{log}\displaystyle{\frac{1}{\sqrt{2\pi |\mathbf{Q}|}}} - \displaystyle{ \frac{1}{2} } (\mathbf{z}-\mathbf{X}\boldsymbol{\beta})^{\top}\mathbf{Q}^{-1}(\mathbf{z}-\mathbf{X}\boldsymbol{\beta}) \label{aln:argmax_b}
\end{align}
If directly apply analytical method to solve the Equation (\ref{aln:argmax_b}) above, then:
\begin{align}
\frac{\partial \operatorname{log}L}{\partial \boldsymbol{\beta}} &= \frac{\partial }{\partial \boldsymbol{\beta}} \left( - \displaystyle{ \frac{1}{2} } (\mathbf{z}-\mathbf{X}\boldsymbol{\beta})^{\top}\mathbf{Q}^{-1}(\mathbf{z}-\mathbf{X}\boldsymbol{\beta}) \right) \nonumber \\
\frac{\partial }{\partial \boldsymbol{\beta}}(\mathbf{z}-\mathbf{X}\boldsymbol{\beta})^{\top}\mathbf{Q}^{-1}(\mathbf{z}-\mathbf{X}\boldsymbol{\beta}) &= \frac{\partial }{\partial \boldsymbol{\beta}} (\mathbf{z}^{\top}\mathbf{Q}^{-1}\mathbf{z} - \mathbf{z}^{\top}\mathbf{Q}^{-1}\mathbf{X}\boldsymbol{\beta} - \boldsymbol{\beta}^{\top}\mathbf{X}^{\top }\mathbf{Q}^{-1}\mathbf{z} + \boldsymbol{\beta}^{\top}\mathbf{X}^{\top }\mathbf{Q}^{-1}\mathbf{X}\boldsymbol{\beta}) \nonumber \\
 &= - \mathbf{z}^{\top}\mathbf{Q}^{-1}\mathbf{X} - \mathbf{X}^{\top }\mathbf{Q}^{-1}\mathbf{z} + \mathbf{X}^{\top }\mathbf{Q}^{-1}\mathbf{X}\boldsymbol{\beta} \label{aln:dLdb}
\end{align}
Set Equation (\ref{aln:dLdb}) as 0, then:
\begin{align}
- \mathbf{z}^{\top}\mathbf{Q}^{-1}\mathbf{X} - \mathbf{X}^{\top }\mathbf{Q}^{-1}\mathbf{z} + \mathbf{X}^{\top }\mathbf{Q}^{-1}\mathbf{X}\boldsymbol{\beta} &= 0 \nonumber \\
\hat{\boldsymbol{\beta}} &= \left(\mathbf{X}^{\top}\mathbf{Q}^{-1}\mathbf{X} \right)^{-1}\mathbf{X}^{\top}\mathbf{Q}^{-1}\mathbf{R}  \nonumber
\end{align}
Second, estimate parameter $\boldsymbol{\rho}$:
\begin{align}
\boldsymbol{\rho}^{(t+1)} &= \underset{\boldsymbol{\rho}}{\operatorname{arg\,max}}\ Q(\boldsymbol{\phi} | \boldsymbol{\phi}^{(t)} ) \nonumber
\end{align}
Assume $\boldsymbol{\rho}=\{ \rho_{1},...,\rho_{k} \}$, without losing any generalizabiliy, $\rho_{1}$ can be estimated as:
\begin{align}
\rho_{1}^{(t+1)} &= \underset{\rho_{1}}{\operatorname{arg\,max}}\ Q(\boldsymbol{\phi} | \boldsymbol{\phi}^{(t)} ) \nonumber
\end{align}
\begin{align}
\frac{\partial \operatorname{log}L}{\partial \rho_{1}} &= \frac{\partial }{\partial \rho_{1}} \left( -\frac{1}{2}\operatorname{log}|\mathbf{Q}| - \displaystyle{ \frac{1}{2} } (\mathbf{z}-\mathbf{X}\boldsymbol{\beta})^{\top}\mathbf{Q}^{-1}(\mathbf{z}-\mathbf{X}\boldsymbol{\beta}) \right) \nonumber \\
\frac{\partial }{\partial \rho_{1}} \operatorname{log}|\mathbf{Q}| &= - \operatorname{tr}(\mathbf{W}_{1}\mathbf{Q}^{-1}) \nonumber \\
\frac{\partial }{\partial \rho_{1}}(\mathbf{z}-\mathbf{X}\boldsymbol{\beta})^{\top}\mathbf{Q}^{-1}(\mathbf{z}-\mathbf{X}\boldsymbol{\beta}) &= \frac{\partial }{\partial \rho_{1}} (\mathbf{z}^{\top}\mathbf{Q}^{-1}\mathbf{z} - \mathbf{z}^{\top}\mathbf{Q}^{-1}\mathbf{X}\boldsymbol{\beta} - \boldsymbol{\beta}^{\top}\mathbf{X}^{\top }\mathbf{Q}^{-1}\mathbf{z} + \boldsymbol{\beta}^{\top}\mathbf{X}^{\top }\mathbf{Q}^{-1}\mathbf{X}\boldsymbol{\beta}) \nonumber
\end{align}
It is impossible to get the analytical solution for $\rho_{i}$. \\

%
%\vspace{-5mm}
Third, estimate parameter $\sigma^{2}$. Let $\sigma^{2}=[\sigma^2]$
\begin{align}
[\sigma^2]^{(t+1)} &= \underset{[\sigma^2]}{\operatorname{arg\,max}}\ Q(\boldsymbol{\phi} | \boldsymbol{\phi}^{(t)} ) \nonumber \\
\frac{\partial \operatorname{log}L}{\partial [\sigma^2]} &= \frac{\partial }{\partial [\sigma^2]} \left( - \frac{1}{2}\operatorname{log}|\mathbf{Q}| - \displaystyle{ \frac{1}{2} } (\mathbf{z}-\mathbf{X}\boldsymbol{\beta})^{\top}\mathbf{Q}^{-1}(\mathbf{z}-\mathbf{X}\boldsymbol{\beta}) \right)
\end{align}
The first term at the the right hand side of equation above is:
\begin{align}
\frac{\partial }{\partial [\sigma^2]} \operatorname{log}|\mathbf{Q}| &= \frac{\partial }{\partial [\sigma^2]} \operatorname{log} \left| I_{n} + [\sigma^2] \left( I_{n} - \sum_{i=1}^{k} \rho_{i} \mathbf{W}_{i} \right)^{-1}\left(\left( I_{n} - \sum_{i=1}^{k} \rho_{i} \mathbf{W}_{i} \right)^{-1}\right)^{\top} \right| \nonumber
\end{align}
The second term is:
\begin{align}
& \frac{\partial }{\partial [\sigma^2]}(\mathbf{z}-\mathbf{X}\boldsymbol{\beta})^{\top}\mathbf{Q}^{-1}(\mathbf{z}-\mathbf{X}\boldsymbol{\beta}) \nonumber \\
&= \frac{\partial }{\partial [\sigma^2]} (\mathbf{z}-\mathbf{X}\boldsymbol{\beta})^{\top} \left( I_{n} + [\sigma^2] \left( I_{n} - \sum_{i=1}^{k} \rho_{i} W_{i} \right)^{-1}\left(\left( I_{n} - \sum_{i=1}^{k} \rho_{i} W_{i} \right)^{-1}\right)^{\top} \right)^{-1} (\mathbf{z}-\mathbf{X}\boldsymbol{\beta}) \nonumber
\end{align}
This is again not solvable by using analytical method.
\subsection{Markov Chain Monte Carlo estimation}\label{apx:MCMC}
The Markov Chain Monte Carlo method generates \x{a sequence of draws that approaches the posterior distribution of interest.} Our solution consists of steps as follows.\\

Step 1. Generate $\mathbf{z}$, $\mathbf{z}$ follows truncated normal distribution. %
\begin{align}
\mathbf{z} \sim \operatorname{TrunNormal}_{n}(\mathbf{X}\boldsymbol{\beta} + \boldsymbol{\theta}, I_{n}) \nonumber
\end{align}
where $I_{n}$ is the $n \times n$ identity matrix. If $y_{i}=1$, then $z_{i} \geq 0$, if $y_{i}=0$, then $z_{i} < 0$ \\

Step 2. Generate $\boldsymbol{\beta}$, $\boldsymbol{\beta} \sim \operatorname{Normal}(\boldsymbol{\nu}_{\beta}, \boldsymbol{\Omega}_{\beta})$
\begin{enumerate}
%2.1
  \item define $\boldsymbol{\beta}_{0}$, where
\[
\boldsymbol{\beta}_{0} = \left[
    \begin{array}{c}
0 \\
0 \\
\vdots \\
0
    \end{array}
    \right]
\]
    %2.2
  \item define $\mathbf{D} = hI_{n}$, $\mathbf{D}$ is a baseline variance matrix, corresponding to the prior $p(\boldsymbol{\beta})$, where $h$ is a large constant, \emph{e.g.} 400.
\[
\mathbf{D}^{-1}
= \left[
    \begin{array}{cccc}
\sigma_{0}^{2} & 0 & \ldots & 0 \\
0 & \sigma_{0}^{2} & \ldots & 0 \\
\vdots & \vdots & \ldots & \vdots\\
0 & 0 & \ldots & \sigma_{0}^{2}
    \end{array}
    \right]
\]
Set $\sigma_{0}^{2}$ as $\displaystyle{\frac{1}{400}}$, a small number close to 0, compared with $\operatorname{Normal}(0,1)$, where $\sigma_{0}^{2}=1$
%2.3
  \item $\boldsymbol{\Omega}_{\beta} = \left(\mathbf{D}^{-1}+ \mathbf{X}^{\top}\mathbf{X}  \right)^{-1}$ \\
        This is because:
        \begin{align}
        \mathbf{z} &= \mathbf{X}\boldsymbol{\beta} + \boldsymbol{\theta} + \boldsymbol{\epsilon} \nonumber \\
        \boldsymbol{\beta} &= \mathbf{X}^{-1}(\mathbf{z}- \boldsymbol{\theta} - \boldsymbol{\epsilon}) \nonumber
        \end{align}
        $\therefore \boldsymbol{\beta} \sim \operatorname{Normal} \left( \mathbf{X}^{-1}(\mathbf{z}-\boldsymbol{\theta}),\
(\mathbf{X}^{\top}\mathbf{X})^{-1} \right)$ \\
        Based on law of initial values, $\boldsymbol{\Omega}_{\beta} = \left( \mathbf{D}^{-1}+\mathbf{X}^{\top}\mathbf{X} \right)^{-1}$
%2.4
  \item Then $\boldsymbol{\nu}_{\beta}$ can be represented by $\boldsymbol{\nu}_{\beta} = \boldsymbol{\Omega}_{\beta}\left(\mathbf{X}^{\top}(\mathbf{z}-\boldsymbol{\theta})+ \mathbf{D}^{-1} \right)$
\end{enumerate}
Step 3. Generate $\boldsymbol{\theta}$, $\boldsymbol{\theta} \sim \operatorname{Normal}(\boldsymbol{\nu}_{\theta}, \boldsymbol{\Omega}_{\theta})$
\begin{enumerate}
%3.1
  \item First, define $\mathbf{B}=I_{n} - \displaystyle{\sum_{i}} \rho_{i} \mathbf{W}_{i} $
  \begin{align}
  \boldsymbol{\theta} &= \displaystyle{\sum_{i}} \rho_{i} \mathbf{W}_{i} + \mathbf{u} \nonumber \\
  (I_{n} - \displaystyle{\sum_{i}} \rho_{i} \mathbf{W}_{i}) \boldsymbol{\theta} &= \mathbf{u} \nonumber \\
  \mathbf{B} \boldsymbol{\theta} &= \mathbf{u} \nonumber \\
  \boldsymbol{\theta} &= \mathbf{B}^{-1} \mathbf{u} \nonumber
  \end{align}
%3.2
  Let $\operatorname{Var}(\mathbf{u})=\sigma^{2}I_{n}$
  \begin{align}
  \operatorname{Var}(\boldsymbol{\theta}) &= \operatorname{Var}(\mathbf{B}^{-1} \mathbf{u}) \nonumber \\
  &= (\mathbf{B}^{\top}\mathbf{B})^{-1} \sigma^{2}I_{n} \nonumber \\
  &= \left( \displaystyle{\frac{\mathbf{B}^{\top}\mathbf{B}}{\sigma^{2}}} \right)^{-1} \nonumber
  \end{align}
  \item Then $\boldsymbol{\Omega}_\theta = \left( I_{n} + \displaystyle{\frac{\mathbf{B}^{\top}\mathbf{B}}{\sigma^{2}}} \right)^{-1}$
  We then add an offset $I_{n}$ to $\displaystyle{\frac{\mathbf{B}^{\top}\mathbf{B}}{\sigma^{2}}}$. So $\boldsymbol{\Omega}_{\theta} = \left( I_{n} + \displaystyle{\frac{\mathbf{B}^{\top}\mathbf{B}}{\sigma^{2}}} \right)^{-1}$
%
%3.3
  \item $\boldsymbol{\nu}_{\theta} = \boldsymbol{\Omega}_{\theta}(\mathbf{z} - \mathbf{X}\boldsymbol{\beta})$, since $\boldsymbol{\theta} = (\mathbf{z}-\mathbf{X}\boldsymbol{\beta}) - \boldsymbol{\epsilon}$
\end{enumerate}
Step 4. Generate $\sigma^{2}$, $\sigma^{2} \sim \operatorname{InvGamma}(a,b)$
\begin{align}
  a &= s_{0} + \displaystyle{\frac{n}{2}} \nonumber \\
  b &= \displaystyle{\frac{2}{\boldsymbol{\theta}^{\top}\mathbf{B}^{\top}\mathbf{B}\boldsymbol{\theta} + \displaystyle{\frac{2}{q_{0}}}} } \nonumber
\end{align}
where $s_{0}$ and $q_{0}$ are the parameters for the conjugate prior of $\sigma^{2}$, and $n$ is the size of data. \\

Step 5. Finally we generate coefficient for $\mathbf{W}$, $\rho_{i}$, using Metropolis-Hasting sampling with a random walk chain. %
\begin{align}
  \rho_{i}^{new} = \rho_{i}^{old} + \Delta_{i},  \nonumber
\end{align}
where the increment random variable $\Delta_{i} \sim \operatorname{Normal}(\nu_{\Delta}, \Omega_{\Delta})$. \\

The accepting probability $\alpha$ is obtained by:
\begin{align}
\operatorname{min} \left( \displaystyle{ \frac{|\mathbf{B}_{new}| \operatorname{exp} \left(- \displaystyle{\frac{1}{2\sigma^{2}}}\boldsymbol{\theta}^{\top}\mathbf{B}_{new}^{\top}\mathbf{B}_{new}\boldsymbol{\theta} \right)}{|\mathbf{B}_{old}| \operatorname{exp} \left(- \displaystyle{\frac{1}{2\sigma^{2}}} \boldsymbol{\theta}^{\top}\mathbf{B}_{old}^{\top}\mathbf{B}_{old}\boldsymbol{\theta} \right)}  }, 1 \right) \nonumber
\end{align}
%
%
%---------------------------------------------------------------
%---------------------------------------------------------------
\subsection{Validation of Bayesian Software}\label{apx:quantile}
One challenge of Bayesian methods is getting an error-free implementation. Bayesian solutions often have high complexity, and a lack of software causes many researchers to develop their own, greatly increasing the chance of software error; many models are not validated, and many of them have errors and do not return correct estimations. So it is very necessary to confirm that the code returns correct results. The validation of Bayesian software implementations has a short history; we \x{wrote a program} using a standard method, the method of posterior quantiles \cite{Cook06}, to validate our software. This method again is a simulation-based method. The idea is to generate data from the model and verify that the software will properly recover the underlying parameters in a principled way. First, we draw the parameters $\theta$ from its prior distribution $p(\Theta)$, then generate data from distribution $p(y\mid\theta)$. If the software is correctly coded, the quantiles of each true parameter should be uniformly distributed with respect to the algorithm output. For example, the $95\%$ credible interval should contain the true parameter with probability $95\%$. Assume we want to estimate the parameter $\theta$ in Bayesian model $p(\theta\mid y)=p(y\mid \theta)p(\theta)$, where $p(\theta)$ is the prior distribution of $\theta$, $p(y\mid \theta)$ is the distribution of data, and $p(\theta\mid y)$ is the posterior distribution. The estimated quantile can be defined as:
\[ \hat{q}(\theta_{0}) = \hat{P}(\theta < \theta_{0}) = \frac{1}{N}\sum_{i=1}^{N} \mathbb{I}(\theta_{i}<\theta_{0}) \]
where $\theta_{0}$ is the true value drawn from prior distribution; $\hat{\theta}$ is a series of draw from posterior distribution generated by the software to-be-tested; $N$ is the number of draws in MCMC. The quantile is the probability of posterior sample smaller than the true value, and the estimated quantile is the number of posterior draws generated by software smaller than the true value. If the software is correctly coded, then the quantile distribution for parameter $\theta$, $\hat{q}(\theta_{0})$ should approaches $\operatorname{Uniform}(0,1)$, when $N \rightarrow \infty$ \cite{Cook06}. The whole process up to now is defined as one replication. If run a number of replications, we expect to observe a uniformly distribution $\hat{q}(\theta_{0})$ around $\theta_{0}$, meaning posterior should be randomly distributed around the true value.\\

\begin{figure}[htbp]
  \begin{center}
    \includegraphics[scale=.35]{\imloc 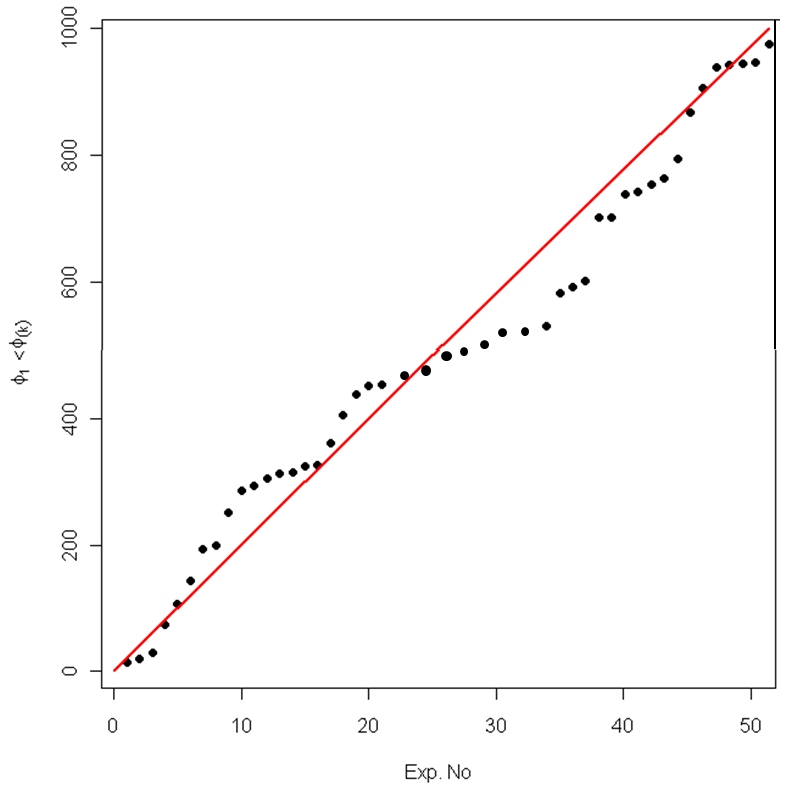}
    \caption{Distribution of sorted quantiles of parameters, $\beta_{1},\beta_{2},\rho_{1},\rho_{2},\sigma^{2}$, over 10 replications. The roughly uniform distribution indicates that the algorithm code functions correctly for data simulated from the model.}\label{fig:quantiles}
  \end{center}
\end{figure}

We then demonstrate the simulations we ran. Assume the model we want to estimate is:
\begin{align}
 \mathbf{z} &= \mathbf{X}_{1}\beta_{1} + \mathbf{X}_{2}\beta_{2} + \boldsymbol{\theta} + \boldsymbol{\epsilon}; \nonumber \\
 \boldsymbol{\theta} &= \rho_{1}\mathbf{W}_{1}\boldsymbol{\theta} + \rho_{2}\mathbf{W}_{2}\boldsymbol{\theta} + \mathbf{u} \nonumber
\end{align}
We then specified a prior distribution for each parameter, and use MCMC to simulate the posterior distributions.
\begin{align}
\boldsymbol{\beta} &\sim \operatorname{Normal}(0, 1); \nonumber \\
\sigma^{2} &\sim \operatorname{InvGamma}(5,10); \nonumber \\
\boldsymbol{\rho} &\sim \operatorname{Normal}(0.05,0.05^{2}) \nonumber
\end{align}
We performed a simulation of 10 replications to validate our hierarchical Bayesian MCMC software. The generated sample size for $\mathbf{X}$ is 50, so the size of the network structure $\mathbf{W}$ is 50 by 50. In each replication we generated 20000 draws from the posterior distribution of all the parameters in $\boldsymbol{\phi}$ ($\boldsymbol{\phi}=\{\beta_{1},\beta_{2},\rho_{1},\rho_{2},\sigma^{2}\}$), and kept one from every 20 draws, yielding 1000 draws for each parameter. We then count the number of draws larger than the true parameters in each replication. If the software is correctly written, each estimated value should be randomly distributed around the true value, so the number of estimates larger than the true value should be uniformly distributed among the 10 replications. We pooled all these quantiles for the five parameters, 50 in total, and the sorted results are shown in Figure \ref{fig:quantiles}.

%
%The X-axis is the total replications of the five parameters -- 50. The Y-axis is the number of draws larger than true parameters in each replication. The red line represents the uniform distribution line. As we can see, the combined results of the five parameters are all uniformly distributed around the true value, thus confirmed that our Bayesian software is correctly written, hence we can apply our software to experiments and return correct estimates.
%
%
%---------------------------------------------------------------
%---------------------------------------------------------------
\subsection{Solution diagnostic}
We run MCMC experiment to confirm there is no autocorrelation among draws of each parameter. In this experiment, we set the length of MCMC chain as 30,000, burn-in as 10,000, and thinning as 20, which is used for removing the autocorrelations between draws. The trace plots \x{generated from our code} for the 1000 draws after burn-in and thinning are listed in the Figure \ref{fig:trace2w} below.

\begin{figure}[htbp]
  \begin{center}
    \includegraphics[scale=.3]{\imloc 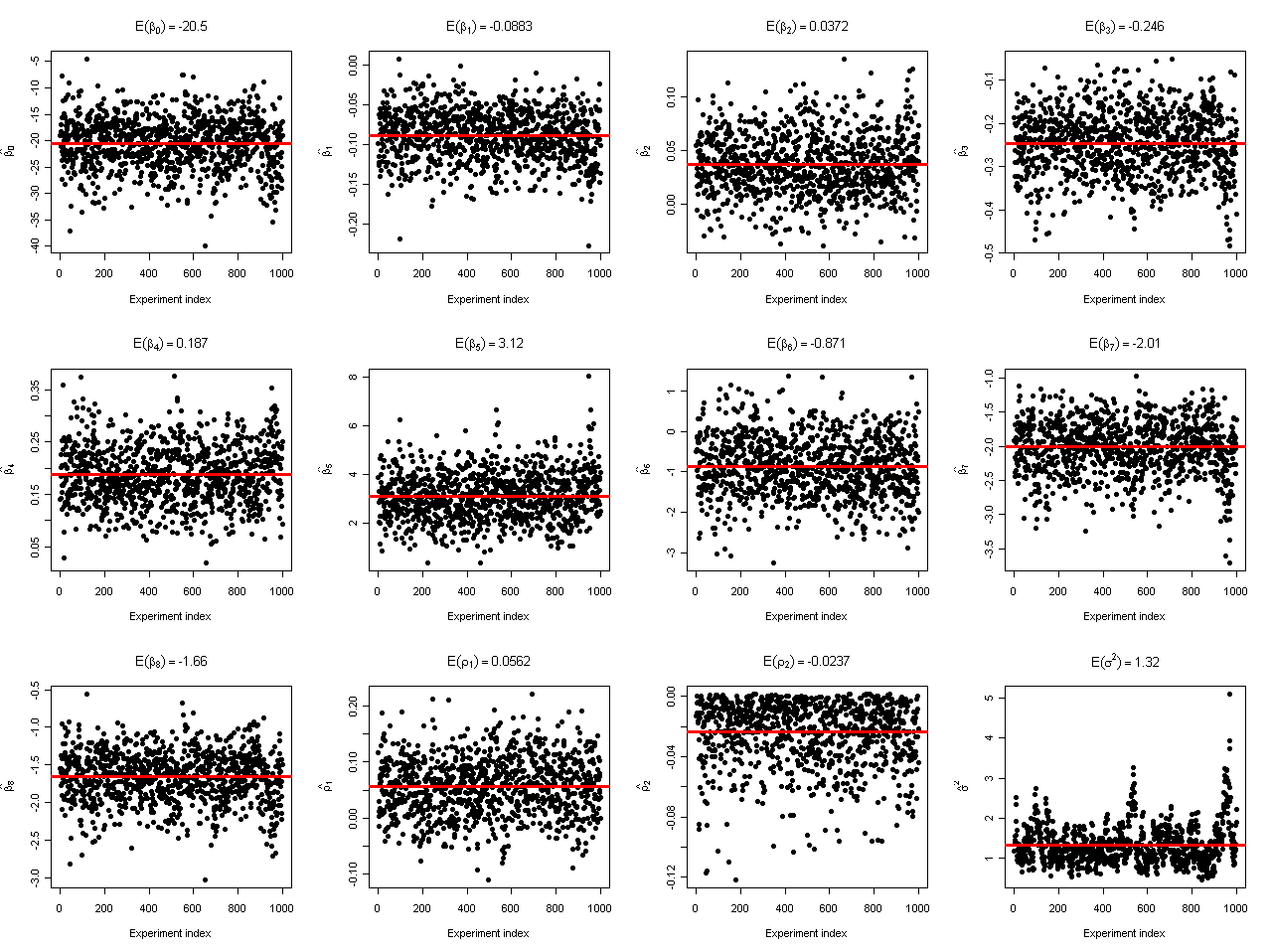}
    \caption{Trace plot of a two-network auto-probit model. \x{$\beta_{0}$: coefficient of constant term, $\beta_{1}$: coefficient of car price; $\beta_{2}$: coefficient of car's optional accessory; $\beta_{3}$: coefficient of consumer's age; $\beta_{4}$: coefficient of consumer's income; $\beta_{5}$: coefficient of consumer's ethnicity; $\beta_{6}$: coefficient of residence longitude; $\beta_{7}$: coefficient of residence latitude; $\rho_{1}$: coefficient of first network autocorrelation term, $\mathbf{W}_{1}$, cohesion; $\rho_{2}$: coefficient of the second network autocorrelation term, $\mathbf{W}_{2}$, structural equivalence; $\sigma^{2}$: estimated variance of the error term in autocorrelation.}}\label{fig:trace2w}
  \end{center}
\end{figure}
We have 12 plots total. Each plot depicts draws for a particular parameter estimation. The first 9 plots, from left to right and top to bottom, are the trace for the $\beta_{i}$, coefficient of independent variables. Each point represents the value of estimated coefficient $\hat{\beta_{i}}$, and the red line represents the mean. We observe all $\hat{\beta_{i}}$s are randomly distributed around the mean, and the mean is significant, showing the estimation results are valid. The 10th and 11th plots are for the two estimated network effect coefficients $\hat{\rho_{1}}$ and $\hat{\rho_{2}}$. We found both $\hat{\rho_{i}}$ are also significant, and randomly distributed around their means. The only coefficient showing autocorrelation is $\sigma^{2}$. \\

\begin{figure}[htbp]
  \begin{center}
    \includegraphics[scale=.45]{\imloc 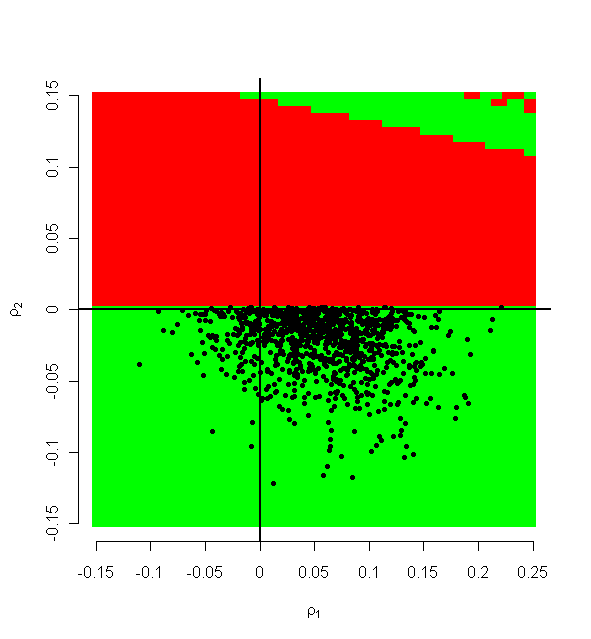}
    \caption{Regions of validity for $\rho_{1}$ and $\rho_{2}$ for which $\mathbf{B}$ is invertible (green) or not (red).}\label{fig:r1r2}
  \end{center}
\end{figure}

Note that not all values of $\rho_{1}$ and $\rho_{2}$ can make $\mathbf{B}$ ($\mathbf{B}=I_{n} - \rho_{1}\mathbf{W}_{1} - \rho_{2}\mathbf{W}_{2}$) invertible. The plot below shows the relationship between the values of $\rho_{1}$ and $\rho_{2}$, and the invertibility of $\mathbf{B}$. The green area is where $\mathbf{B}$ is invertible, and red area is otherwise. If limit draws to the green area, we will have correlated $\rho_{1}$ and $\rho_{2}$. When we draw $\rho_{1}$ and $\rho_{2}$ using bivariate normal, there is no apparent correlation between them (see Figure \ref{fig:r1r2}). We understand the correlation between $\rho_{1}$ and $\rho_{2}$ comes from the definition of $\mathbf{W}_{1}$ and $\mathbf{W}_{2}$, not the prior non-correlation.

%
%---------------------------------------------------------------
%---------------------------------------------------------------
%
\subsection{W as a mixture of matrices}\label{apx:W}
Yang and Allenby \citeyear{Yang03}) specified the autoregressive matrix W as a finite mixture of coefficient matrices, each related to a specific covariate:
\begin{align}
\mathbf{W} = \sum_{i=1}^{n}\phi_{i}\mathbf{W}_{i} \nonumber \\
\sum_{i=1}^{n}\phi_{i} = 1 \nonumber
\end{align}
where $i$ represents the indices of the covariates, $i$ = 1... $n$. $\phi_{i}$ is the correspondent weight of the component matrix $\mathbf{W}_{i}$.  $\mathbf{W}_{i}$ is associated with a covariate $\textbf{X}_{i}$.

%\appendixhead{ZHANG}

% Acknowledgments
%\begin{acks}
%This work was supported in part by AT\&T and the iLab at Heinz College, Carnegie Mellon University.
%\end{acks}

% History dates
%\received{April 2012}

% Electronic Appendix
%\elecappendix

%\medskip

%\section{This is an example of Appendix section head}

\end{document}